# Métodos Empíricos Aplicados à Análise Econômica do Direito

Thomas Victor Conti[1]
2020

**Resumo**

Nas últimas décadas houve forte mudança no perfil das publicações em análise econômica do direito e nos métodos empíricos mais utilizados. Porém, nos próximos anos a mudança pode ser maior e mais rápida, exigindo adaptação dos pesquisadores da área. Neste capítulo analiso algumas tendências recentes de mudança e oportunidades futuras que se avizinham a partir avanços nas bases de dados, estatística, computação e no arcabouço regulatório dos países. Avanço a hipótese de que expansão de objetos e métodos favorecerá equipes de pesquisa maiores e interdisciplinares e apresento evidências circunstanciais a partir de dados bibliométricos de que isso já vem acontecendo no *Journal of Law and Economics*.

**Abstract**

In the last decades there has been a strong change in the profile of publications on economic analysis of law and in its most used empirical methods. However, in the coming years change may be greater and faster, requiring researchers in the area to adapt. In this chapter I analyze some recent trends of change and future opportunities that are coming from advances in databases, statistics, computing and in some countries' regulatory framework. I advance the hypothesis that expansion of objects and methods will increasingly favor larger and interdisciplinary research teams and I present circumstantial evidence using bibliometric data that this is already happening in the *Journal of Law and Economics*.

[1] Professor do Insper e do IDP-SP. Email: thomasvc@insper.edu.br

## 1. Introdução

Mais de 40 anos após o início da chamada *revolução da credibilidade*[2] nos estudos econômicos, ela continua a pleno vapor. Enquanto em 1960 mais da metade dos artigos publicados nos periódicos mais prestigiados de Economia eram puramente teóricos, em 2010 esse percentual havia caído para 19,1%.[3] No mesmo período, artigos empíricos usando dados públicos ou privados aumentaram de 45% para 64,3%, e dois métodos praticamente inexistentes – os trabalhos empíricos usando experimentos e o uso de simulações para testar teorias – em 2010, respondiam somados por 17% do total de artigos publicados nestes periódicos.

Embora números mais atualizados para o biênio 2019/2020 ainda não estejam disponíveis, a tendência consistente indica que a fatia dos estudos empíricos (incluindo experimentos) e simulações deve ter crescido mais um pouco e a teoria pura continuará a perder um pouco de espaço.

Pesquisas de abordagem empírica também cresceram nos periódicos especializados de Direito em língua inglesa. Entre 1990 e 2009 praticamente, duplicaram.[4] Artigos mencionando a palavra "empírico", em inglês, eram 0% das publicações de 1950 e perto de 60% em 2010.[5] Daniel Ho e Larry Kramer defendem que o Direito também passa por uma revolução empírica:

> O fato de que não apenas especialistas e defensores dedicados, mas também uma ampla gama de acadêmicos, cortes e tomadores de decisão estão lidando com dados é um sinal da vitalidade da revolução empírica. Essa tendência pode ser mais visível em Stanford do que em outras escolas de Direito, porém o movimento está por toda parte.[6]

---

[2] ANGRIST, Joshua D.; PISCHKE, Jörn-Steffen. "The Credibility Revolution in Empirical Economics: How Better Research Design Is Taking the Con out of Econometrics". Journal of Economic Perspectives, vol. 24, nº 2 (2010), pp. 3 – 30.
[3] HAMERMESH, Daniel S. "Six Decades of Top Economics Publishing: Who and How?". Journal of Economic Literature, vol. 51, nº 1 (2013), pp. 162 – 172.
[4] HEISE, Michael. "An Empirical Analysis of Empirical Legal Scholarship Production, 1990-2009". University of Illinois Law Review, vol. 2011 (2011), p. 1739.
[5] HO, Daniel E.; KRAMER, Larry. "Introduction: The Empirical Revolution in Law". *Stanford Law Review*, vol. 65, nº 6 (2013), pp. 1195 – 1202.
[6] HO, Daniel E.; KRAMER, Larry, Introduction: The Empirical Revolution in Law, **Stanford Law Review**, v. 65, n. 6, p. 1195–1202, 2013.

Diante desse quadro, estudiosos do Direito têm muito espaço de pesquisa a ganhar, avançando em suas preocupações empíricas e ficando a par dos novos métodos de trabalho e pesquisa com dados. Neste artigo apresento um panorama breve, não exaustivo, com o intuito de ajudar mostrar a relevância dos novos métodos e um pouco do panorama atual em Direito e Economia.

Na Parte II exponho as formas mais utilizadas atualmente de se obter bases de dados adequadas para pesquisas empíricas em Análise Econômica do Direito (AED). Hoje, muito do potencial inovador de uma pesquisa está no ineditismo da base de dados que lhe dá suporte. Destaco a importância crescente das bases de dados privadas e de métodos algorítmicos de extração de dados, e aponto alguns dos seus problemas subjacentes.

Na Parte III comento sobre a grande área da inferência causal, as contribuições e os desafios que trouxe para a AED. Em geral, estudos empíricos têm como motivação ganhar um entendimento sobre uma intervenção, ação ou mudança, que foi ou que será realizada na sociedade. Para isso, é preciso entender as causas reais por trás dos fenômenos, distinguindo-as de meras correlações e as separando de outras variáveis envolvidas. Diferenciar correlação de causalidade é bem mais complexo do que parece à primeira vista.

Na Parte IV destaco outra gama de estudos empíricos na AED cuja importância tem crescido no mundo todo, inclusive no Brasil. São os relatórios técnicos de Análises de Custo-Benefício para reformas jurídicas. Embora pouco presentes na academia, a crescente pressão sobre todos os níveis da administração pública por mais transparência e acompanhamento de resultados têm levado à necessidade de estimar os efeitos de mudanças jurídicas. O arcabouço teórico e metodológico da AED se encontra especialmente bem posicionado para responder a essas demandas.

O capítulo termina com algumas considerações a respeito dos métodos empíricos apresentados, possibilidades de pesquisa e publicação.

**2. Obtenção de dados**

A pesquisa empírica avançou, em parte naturalmente, como resposta ao avanço tecnológico. O custo para armazenar, compartilhar e analisar dados tem se reduzido ano após ano, aumentando a oferta de dados para análise em uma pesquisa acadêmica nas ciências sociais aplicadas.

Dados jurídicos se tornaram objeto especial de interesse de mais de uma área do conhecimento. Propriedade privada, regulação e impacto regulatório, meio ambiente, indenizações, crime, corrupção, contratos, impostos, dentre outros temas, envolvem uma interação constante com os dispositivos legais. Usando métodos contemporâneos de raspagem de dados – números, textos, imagens – é possível compilar bases com processos, decisões judiciais e mudanças jurídicas com alto nível de detalhamento.

Centros de pesquisa e pesquisadores individuais têm montado bases desse tipo que, depois de organizadas e limpas, podem servir para mais de um estudo. Tem sido cada vez mais comum pesquisadores que estruturaram bases de dados de interesse amplo colocarem-nas à disposição pública, via repositório online ou criação de um pacote estatístico para linguagens de programação, como *R* ou *Python*. Junto com a gestão da base de dados ou do código os pesquisadores responsáveis lançam um artigo acadêmico explicando a metodologia por trás da criação e organização da base. Outros pesquisadores que façam uso da base criada – em geral gratuitamente – pede-se que citem o artigo que explica o método da base para dar os devidos créditos aos criadores. Artigos sobre novas bases de dados são cada vez mais bem vindos em periódicos da área temática mais próxima da base de dados, porém quando essa modalidade de publicação ainda não é aceita, há periódicos especializados em publicação de novas bases de dados ou novos pacotes estatísticos para análise de dados.

A sistematização destas iniciativas é um fator determinante para alavancar as pesquisas empíricas no Direito. Identificar as principais fontes de bases de dados jurídicas organizadas, como extraí-los, onde publicar estudos etc., forma parte importante do conhecimento de fronteira para expandir os métodos empíricos.[7]

---

[7] REEVE, Allison C.; WELLER, Travis. "Empirical Legal Research Support Services: A Survey of Academic Law Libraries". *Law Library Journal*, vol. 107 (2015), p. 399.

Um exemplo recente da importância deste tipo de iniciativa alinhada com a abordagem da Análise Econômica do Direito foi a criação do *Coronavirus Government Response Tracker*, da *Blavatnik School of Government* na Universidade de Oxford. A equipe de pesquisadores mapeia com a máxima velocidade possível todas as leis que vêm sendo adotadas pelo mundo na tentativa de proporcionar o distanciamento social. Também criam um indicador da força do isolamento, chamado *Stringency Index*. Os resultados são disponibilizados na forma de gráficos interativos, download da base de dados completa organizada por país e data, e uma API para fácil comunicação com softwares automatizados de pesquisadores do mundo todo.[8] A Figura 1 ilustra o resultado da base de dados criada para alguns países selecionados.

No contexto da pandemia do novo coronavírus é inquestionável a importância de termos conhecimento empírico do Direito aplicado em diferentes lugares, com metodologia sólida que permita a comparação. Embora recente, a iniciativa foi citada em editorial da revista *Nature* destacando a importância de comparações internacionais e de compartilhamento de melhores práticas[9] – algo difícil de realizar sem um indicador que ajude a filtrar as medidas. Métodos empíricos no Direito em geral e na Análise Econômica do Direito apenas generalizam a importância dessa abordagem realista quanto às consequências das normas jurídicas para todas as outras áreas do Direito.

---

[8] HALE, Thomas; WEBSTER, Sam; PETHERICK, Anna *et al. Oxford COVID-19 Government Response Tracker*. Blavatnik School of Government. Oxford, UK, 2020.
[9] NATURE. "Coronavirus: share lessons on lifting lockdowns". *Nature*, vol. 581 (06 mai. 2020), p. 8.

**Figura 1 – Comparação das respostas de seis países (*stringency index*) conforme aumentaram os casos de coronavírus**

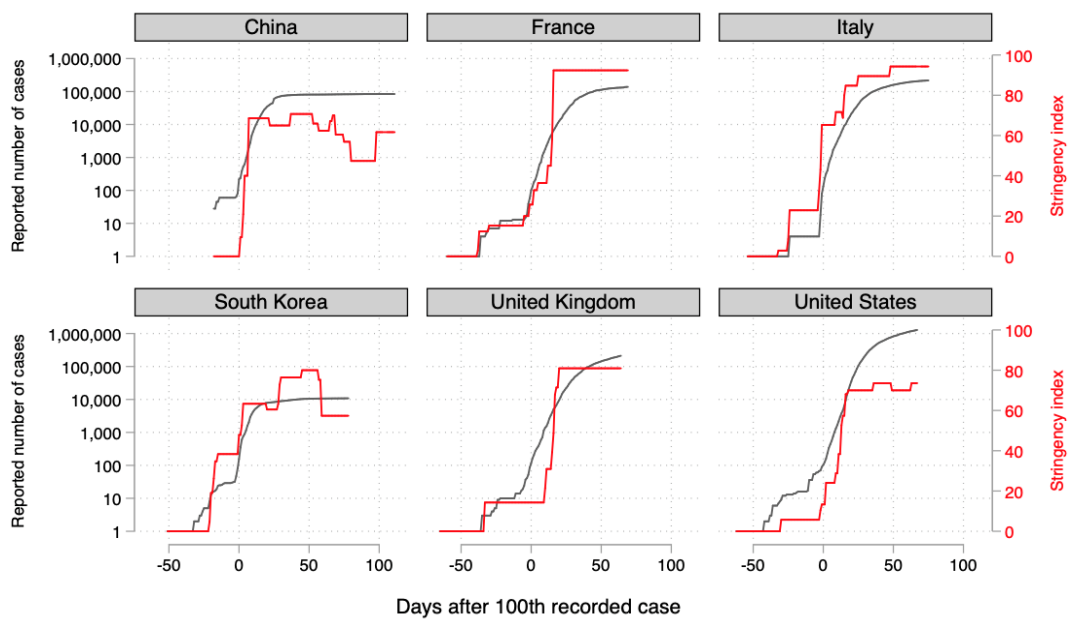

Dados de 9 de maio de 2020. Dados de países individuais podem estar com atraso.
Fonte: Oxford COVID-19 Government Response Tracker.[10]

No contexto da pandemia do novo coronavírus é inquestionável a importância de termos conhecimento empírico do Direito aplicado em diferentes lugares, com metodologia sólida que permita a comparação. Embora recente, a iniciativa foi citada em editorial da revista *Nature* destacando a importância de comparações internacionais e de compartilhamento de melhores práticas[11] – algo difícil de realizar sem um indicador que ajude a filtrar as medidas. Métodos empíricos no Direito em geral e na Análise Econômica do Direito apenas generalizam a importância dessa abordagem realista quanto às consequências das normas jurídicas para todas as outras áreas do Direito.

Exemplos de extração/obtenção e manejo de dados como este resultam da melhor tecnologia de dados e redes organizadas de pesquisa pelo mundo, mas não chegam a fazer uso de técnicas estatísticas inovadoras na criação da própria base de dados. Dois métodos mais recentes trazem um potencial promissor de avançar na expansão da fronteira de possibilidades de análises empíricas em Análise Econômica do Direito.

---

[10] *Ibid.*
[11] NATURE. "Coronavirus: share lessons on lifting lockdowns". *Nature*, vol. 581 (06 mai. 2020), p. 8.

Um deles é o uso de técnicas de aprendizado de máquina (*machine learning*) para mapear quantidades imensas de texto escrito, identificar padrões ou realizar sínteses que permitam estudos empíricos radicalmente inovadores. Criação de redes de legislação,[12] redes neurais para resumir documentos,[13] e/ou processamento de linguagem natural para identificar temas mais relevantes em determinado cenário ou filtrar uma base proibitivamente grande de ser analisada.[14] Por enquanto esta modalidade de aplicações tem sido mais comumente encontrada em empresas *startups* voltadas a oferecer produtos para o mercado jurídico. No entanto, dada a falta de profissionais qualificados no mercado e a crescente competitividade nas publicações acadêmicas, é de se esperar que é apenas uma questão de tempo até que métodos similares venham a revolucionar a forma com que a pesquisa empírica no Direito e Economia é feita.

O outro método promissor de fronteira para viabilizar novas modalidades de bancos de dados e pesquisas empíricas é o chamado *data-fusion*, proposto por estudos do renomado cientista da computação Judea Pearl. Trata-se do uso de técnicas estatísticas avançadas para permitir o uso simultâneo de bancos de dados diferentes, coletados de forma distinta e com potenciais vieses múltiplos, fundindo-os em um único banco de dados que permita fazer inferências e estudos empíricos específicos, bem desenhados. Pode permitir um tratamento mais rigoroso do problema da generalização ou validade externa dos modelos estimados na análise empírica.[15] O potencial desta metodologia ainda não foi aproveitado, porém os avanços teóricos – todos bastante recentes – têm se desenvolvido e também é possível que no futuro tenhamos um aumento na adoção desse tipo de técnica por praticantes.

---

[12] SAKHAEE, Neda; WILSON, Mark C. "Information extraction framework to build legislation network". *Artificial Intelligence and Law*, 2020.
[13] TRAN, Vu; LE NGUYEN, Minh; TOJO, Satoshi *et al*. "Encoded summarization: summarizing documents into continuous vector space for legal case retrieval". *Artificial Intelligence and Law*, 2020.
[14] DALE, Robert. "Law and Word Order: NLP in Legal Tech". *Natural Language Engineering*, vol. 25, nº 1 (2019), pp. 211 – 217; MOK, Wai Yin; MOK, Jonathan R. "Legal Machine-Learning Analysis: First Steps towards A.I. Assisted Legal Research". *In Proceedings of the Seventeenth International Conference on Artificial Intelligence and Law.* Montreal: Association for Computing Machinery, 2019, pp. 266 – 267.
[15] BAREINBOIM, Elias; PEARL, Judea. "Causal inference and the data-fusion problem". *Proceedings of the National Academy of Sciences*, vol. 113, nº 27 (2016), pp. 7345 – 7352.

A dinâmica de crescente trabalho para criação de bases de dados novas e mais sofisticadas, somada à importância crescente da análise empírica para publicação nos melhores periódicos, pode levar a problemas científicos e éticos na condução da pesquisa. A possibilidade de uma pesquisa com dados inéditos é um incentivo para que acadêmicos desejem acessar microdados de grandes empresas, podendo gerar conflito de interesse entre o resultado da pesquisa e empresas que também queiram usar uma publicação acadêmica como estratégia publicitária.[16]

## 3. Inferência Causal

O lema "correlação não é causalidade" é comumente repetido em faculdades de Economia e por todas as áreas, porém raramente há um avanço em relação a essa afirmação correta. Se causalidade não é correlação, o que é a causalidade e como identificá-la? [17] A grande área da inferência causal pode ser resumida como uma tentativa de criar métodos que sejam consistentes na resposta a essa pergunta.

O raciocínio da causalidade exige, pelo menos, a noção de que uma mudança em uma variável (causa) levará a uma mudança em outra variável (efeito), ainda que entre as causas e os efeitos existam uma série de variáveis mediadoras difíceis de identificar, e a precedência temporal das causas ante os efeitos seja de prazos longos que também dificultem a observação. Além disso, avaliar a existência de uma relação causal exige também um raciocínio lógico abdutivo.[18] Na abdução é necessário raciocinar considerando causas potenciais alternativas e efeitos potenciais alternativos e avaliar, por contrafactuais, o que ocorreria na ausência de uma ou mais destas causas, se o resultado se manteria ou não, e com qual relevância.

---

16 Um exemplo recente do potencial para esse tipo de conflito é a crítica pelo economista Hubert Horan a algumas pesquisas econômicas sobre a *Uber*, feitas com bases de dados inacessíveis da empresa a outros pesquisadores, e que se tornaram matérias em jornais ainda em preprint e com potenciais conflitos entre autores e a empresa. HORAN, Hubert. *Uber's "Academic Research" Program: How to Use Famous Economists to Spread Corporate Narratives*. Promarket. 5 dez. 2019; Matéria e investigação do *Times Higher Education* também encontrou conflitos de interesse nessa agenda de pesquisa e expressou preocupação. MATTHEWS, David. *Bias fears over Uber academic research programme*. Times Higher Education. ; Discussões mais gerais sobre ética profissional e de publicação em Economia pode ser encontrada em DEMARTINO, George; MCCLOSKEY, Deirdre N. *The Oxford Handbook of Professional Economic Ethics*. Oxford: Oxford University Press, 2016.
[17] PEARL, Judea; MACKENZIE, Dana. *The Book of Why*: The New Science of Cause and Effect. 1ª ed. New York: Basic Books, 2018.

Desta forma, a causalidade é mais exigente que a correlação. Por exemplo, caso se espere que uma nova lei ambiental reduza a poluição, não é suficiente observar que após a aprovação da lei o nível de poluição caiu para concluir que o efeito foi produzido pela nova lei. Esta associação seria no máximo uma correlação negativa entre a força do dispositivo legal e o nível de poluição – quanto mais forte a lei, menor a poluição. O raciocínio causal exige que encontremos evidências suficientes para afirmar que *na ausência da lei o nível de poluição teria sido maior*. Isto é, a relação causal entre os incentivos criados pela lei e o nível final de poluição permite, inclusive, que após a adoção da lei o nível de poluição aumente, desde que tenhamos boas razões e evidências para acreditar que na ausência da lei a poluição teria *aumentado mais*.

Nosso raciocínio causal funciona bem para atividades cotidianas. Em situações sociais mais complexas e/ou mais dispersas ao longo do tempo, ele pode, e costuma falhar, e nesses casos dependemos do rigor teórico e empírico para pisar em chão firme. No entanto, encontrar evidências do que poderia ter acontecido é muito difícil, quando não impossível, pois é necessário avaliar e excluir toda sorte de possíveis outras causas e efeitos que não foram observados ou não são conhecidos/imaginados – é o chamado viés de variável omitida.

Nas ciências físicas e biológicas, o contexto de laboratório de experimentos permite uma engenharia de controle destas muitas variáveis externas alternativas. Um exemplo extremo do nível de controle atingido na física é o caso do sensor LIGO, que para detectar a existência de ondas gravitacionais necessitou de precisão para captar uma perturbação 10 mil vezes menor que o raio de um próton.[19] Só foi possível certificar que uma perturbação dessa magnitude no sensor seria de fato causada por uma onda gravitacional após excluir todas as outras fontes possíveis de perturbação, com câmaras à vácuo gigantescas e até ajustes para corrigir o efeito da curvatura da Terra sobre os sensores.

---

[18] DOUVEN, Igor. "Abduction". *In* ZALTA, Edward N. (ed.). *The Stanford Encyclopedia of Philosophy*. Stanford University, 2017.

É evidente que este nível de precisão e controle sobre variáveis externas é inalcançável nas ciências sociais aplicadas. Quando não impossíveis *de facto*, são impossíveis *de jure* – direitos fundamentais e a ética impedem que se realizem experimentos tão intrusivos. Apesar das limitações epistêmicas impostas por essas condições materiais e jurídicas, ter um maior grau de confiança sobre a resposta para perguntas causais é uma necessidade política e social.

No limite, toda lei, regulação e decisão judicial pressupõe algum modelo causal a respeito dos efeitos que produzirão na realidade. Abandonar a busca de respostas científicas para o desafio da causalidade muitas vezes é o mesmo que aceitar respostas muito ruins, modelos inconsistentes e desprovidos de qualquer evidência.

A dificuldade de lidar com a causalidade leva a tomar como padrão ouro da evidência científica os experimentos aleatórios controlados (*Randomized Controlled Trials – RCTs*). Em RCTs, embora raramente se tenha um nível de controle como em um laboratório, é central a preocupação de criar grupos comparáveis e aplicar um tratamento em apenas um deles. Este tratamento, se distribuído aleatoriamente, não tem outras causas. Assim, a diferença observada no efeito médio de interesse entre o grupo tratado e o não tratado poderá ser interpretada como uma diferença causada pelo tratamento – satisfeitos os diversos pressupostos do experimento.

Em Direito e Economia muitas vezes esta separação de grupos se dá por limites municipais ou estaduais entre quem adotou ou não adotou determinada legislação. Em outros casos como em Economia comportamental pode-se tornar aleatório o envio de determinado *nudge* para parte de uma população e não outra. Políticas públicas podem passar por uma etapa avaliativa de menor escala na qual a separação de grupos ajuda a interpretar os efeitos causais.

---

[19] LIGO - LASER INTERFEROMETER GRAVITATIONAL-WAVE OBSERVATORY. *Quick Facts about LIGO*. LIGO Lab Caltech.

O avanço das pesquisas econômicas usando desenhos de experimentos aleatórios controlados se deu a partir dos anos 1990.[20] Em 2019, os economistas Abhijit Banerjee, Esther Duflo e Michael Kremer foram laureados com o Prêmio Nobel de Economia "pela abordagem experimental em reduzir a pobreza global",[21] sendo grandes expoentes deste programa de pesquisa focado em estudos controlados.

No entanto, embora promissora, a estratégia dos experimentos controlados possui muitos limites práticos. Gelbach e Klick destacam que, praticamente toda a área de Análise Econômica do Direito Penal ou Processual Penal não permitiria conduzir estudos desse tipo por questões éticas.[22] Em outros casos o limite se daria por impossibilidade financeira de custear grupos de controle grandes ou experimentos conduzidos por prazos muito longos.

Há um movimento recente, controverso e interno à área do Direito que defende que os benefícios potenciais de uma distribuição aleatória em pequena escala de uma lei para sabermos seus efeitos causais são grandes e os custos pequenos,e que a flexibilização de algumas destas regras pode ser benéfica para a pesquisa e melhoria da legislação.[23] Iniciativas muito inovadoras à parte, segue verdadeiro que diversas situações sociais não poderão ser estudadas com desenho de pesquisa RCT, o que leva a busca dos melhores substitutos possíveis.

Os desenhos experimentais e quase experimentais foram adotados desde cedo na Análise Econômica do Direito. As mudanças legais em diferentes níveis de jurisdição, próximos uns dos outros, permitem que grupos próximos sejam afetados por leis muito diferentes e as comparações sejam interpretadas como potencialmente causais. Como o experimento nunca é perfeito, métodos

---

estatísticos diversos foram criados para separar a variabilidade gerada por variáveis omitidas nos dados ou no modelo, mas conhecidas na prática por conhecimento intuitivo.

Métodos como Diferenças em Diferenças, Variáveis Instrumentais, Controle Sintético, *Propensity Score*, dentre outros, são adotados na literatura de fronteira da AED levando a melhorias substanciais na qualidade da pesquisa empírica, formulação de novas hipóteses e teorias. Cada método possui um conjunto de pressupostos necessários para que as estimativas encontradas para o efeito tenham, de fato, interpretação causal, e alguns desses pressupostos precisam ser assumidos como verdadeiros à luz apenas da intuição ou conhecimento especializado da área – fato "inevitável e desconfortável"[24].

Apesar do desconforto há também oportunidades nos estudos observacionais que podem ser perdidas com experimentos aleatórios de campo ou em laboratório. Em ciências sociais o processo de tornar aleatório, ou a moldura laboratorial necessária para estes estudos, podem contaminar os resultados com outras variáveis omitidas que não existiriam em um estudo observacional. Um estudo laboratorial experimental que tenha identificado um efeito causal com precisão precisa lidar com o problema de não saber se o efeito encontrado é generalizável para outro contexto. Um estudo observacional tem maior dificuldade de identificar o efeito causal, porém se o fizer de forma competente pode ter maior confiança nessa generalização ao menos para o contexto regional/jurisdicional que foi estudado.

Esta temática da inferência causal tem passado por grandes inovações teóricas e metodológicas. Duas grandes abordagens para interpretação de modelos causais são a de Neyman-Rubin[25] (também conhecido como *potential outcomes*) e a de Judea Pearl[26] dos Grafos Direcionados Acíclicos (*Directed Acyclic Graphs – DAGs*).

---

[24] GELBACH; KLICK, *op. Cit.*
[25] IMBENS, Guido W.; RUBIN, Donald B. *Causal Inference for Statistics, Social, and Biomedical Sciences: An Introduction.* New York: Cambridge University Press, 2015.
[26] PEARL, Judea. *Causality*: Models, Reasoning and Inference. 2ª ed. New York: Cambridge University Press, 2009.

Em AED e em Economia, como um todo, a abordagem Neyman-Rubin é muito mais tradicional em influente. Mas a formalização de Pearl para os efeitos causais apoiados em DAGs foi provada, matematicamente completa, apenas recentemente,[27] e tem havido uma profusão de novos avanços metodológicos usando o arcabouço matemático criado. Algoritmos novos para modelar relações causais extremamente complexas, ou até para inferir um modelo causal a partir dos dados, são inovações instigantes.

Guido Imbens, que desenvolveu com Angrist e Rubin o método de variáveis instrumentais e outras técnicas indispensáveis de inferências causais, recentemente publicou um longo texto para discussão comparando as abordagens de Pearl e Neyman-Rubin, concluindo que a abordagem de Pearl "merece a atenção de todos os pesquisadores e usuários da inferência causal como uma das suas metodologias principais". Imbens credita a não adoção de DAGs à falta de estudos empíricos em Economia, até o momento, que demonstrem com força e convencimento o potencial do método.[28] Assim, embora pouco representado na literatura de AED e com prováveis dificuldades de aceitação em periódicos de Economia comparado com o arcabouço dos POs (*potential outcomes*), na humilde opinião deste autor vale a pena acompanhar o desenvolvimento de novos estudos empíricos nessa área, pois há um latente potencial inovador.

Um último ponto relevante sobre inferência causal e os métodos supracitados é que cada vez menos a aplicação deles em temas do Direito ou da AED está condicionada à pesquisa nestas áreas. Estatísticos, cientistas da computação, epidemiologistas, formuladores de políticas públicas e outros especialistas de áreas distintas têm cada vez mais adotado métodos empíricos tendo leis, legislação e decisões judiciais como objeto de pesquisa.

Um exemplo representativo desse fenômeno é a pesquisa deste momento, de crise mundial do coronavírus, sobre os impactos de diferentes leis de restrição de mobilidade ou de normas sanitárias sobre a velocidade de contágio do vírus. Embora os epidemiologistas sejam por excelência os mais habilitados para prever o comportamento do vírus e da epidemia, o impacto das leis de restrição sobre o comportamento das pessoas seria por excelência a área de estudo da Análise Econômica do Direito. No entanto, pesquisadores de todas as áreas estão interessados no tema e farão pesquisas sobre isso.

---

[27] HUANG, Yimin; VALTORTA, Marco. "Identifiability in causal Bayesian networks: a sound and complete algorithm". *In Proceedings of the 21st national conference on Artificial intelligence - Volume 2*. Boston, Massachusetts: AAAI Press, 2006, pp. 1149–1154.

[28] IMBENS, Guido W., Potential Outcome and Directed Acyclic Graph Approaches to Causality: Relevance for Empirical Practice in Economics, **Journal of Economic Literature**.

A generalização dos métodos estatísticos avançados para todas as áreas da ciência permite essa diversidade, ficando o avanço teórico econômico comportamental como o grande diferencial da AED para se distinguir destas outras abordagens. [29] A temática geral ganha escala e melhores contribuições por um lado, porém pressiona os pesquisadores do Direito e da AED a uma constante necessidade de atualização. Não é fácil assimilaras técnicas de inferência na fronteira do conhecimento estatístico, de um lado, e o olhar abrangente para periódicos e pesquisas de muitas outras áreas para acompanhar a fronteira do conhecimento no tema de interesse.

Diante dessa pressão acredito que a pesquisa empírica em AED será realizada cada vez mais por grupos de pesquisa, e provavelmente, com grupos de pesquisa cada vez mais interdisciplinares.

Apenas por curiosidade, explorei essa hipótese extraindo metadados por *webscraping* dos artigos publicados no *Journal of Law and Economics* (JLE) de 1957 a 2019, totalizando 1.600 artigos. Por limitações no formato dos artigos e disponibilização não gratuita, ative-me aos dados dos artigos disponibilizados com Resumo (*Abstract*) público e com caracteres reconhecíveis, ficando apenas os dados de 697 artigos publicados de 1993 a 2019. A Figura 2 abaixo mostra o crescimento no número médio de autores por artigos no JLE:

---

[29] PANHANS, Matthew T.; SINGLETON, John D. "The Empirical Economist's Toolkit: From Models to Methods". *History of Political Economy*, vol. 49, nº suplementar (2017), pp. 127 – 157.

**Figura 2 – Número médio de autores por artigo publicado no *Journal of Law and Economics*, 1994-2019, média móvel de 3 anos**

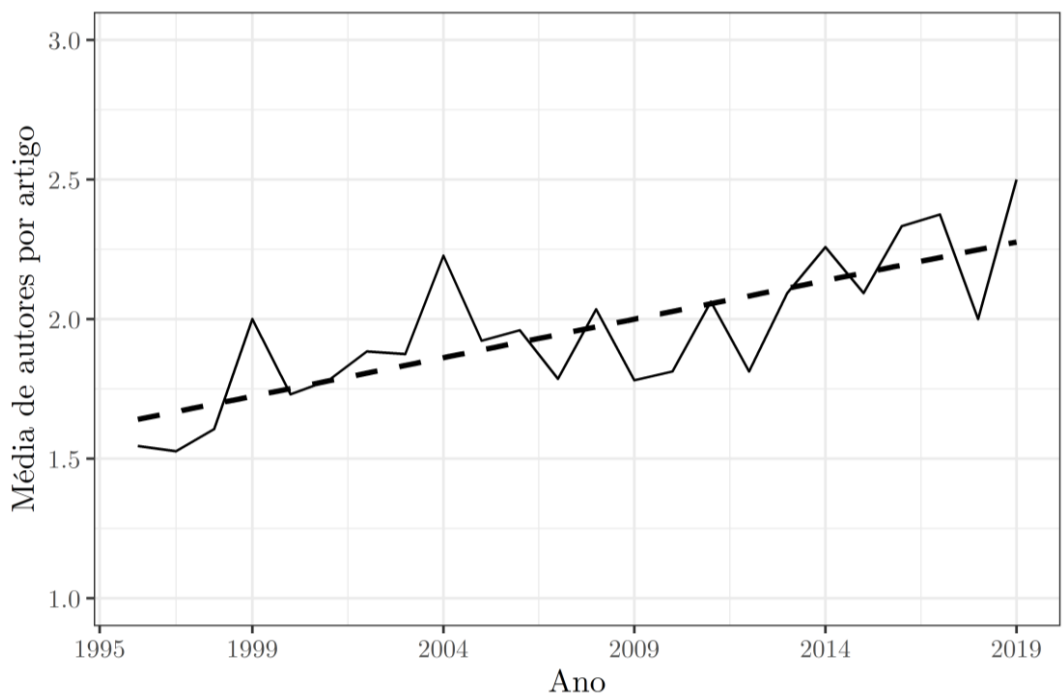

Fonte: Elaboração própria a partir de dados públicos extraídos do *Journal of Law and Economics*.

Vemos que o número médio de autores passou de 1,5 em torno de 1993-1995 para 2,5 por artigo em 2019. Esse número certamente é afetado por muitas outras variáveis que não apenas a pressão pela fronteira empírica, como por exemplo, os próprios incentivos à publicação e número de citações. Porém é razoável supor que as competências diversas necessárias para (1) conhecer o fenômeno jurídico e econômico de fundo; (2) conhecer a fronteira das técnicas estatísticas mais adequadas para o caso em questão; e (3) programar a extração de dados ou a organização de bases de dados cada vez mais complexas, são difíceis de se encontrar em uma única pessoa sem que haja prejuízos para a eficiência do desempenho em alguma dessas funções.

A mensagem é otimista para todos os interessados na área da Análise Econômica do Direito que se sintam impedidos por não estarem na fronteira estatística ou computacional – poucos estão. Minha hipótese é que o futuro da AED no Brasil e no mundo estará, em grande medida, dependente da criação de grupos de pesquisa interdisciplinares com forte espírito colaborativo e disposição a criar pontes de conhecimento.

## 4. Análise Custo-Benefício e Previsão

Uma linha diferente de método empírico em Análise Econômica do Direito busca trazer respostas para problemas que exigem previsão de resultados. Trata-se da Análise Custo-Benefício(*Cost Benefit Analysis – CBA*), na qual o objetivo central é a avaliação dos benefícios potenciais *versus* custos de oportunidade de diferentes leis, regulações, decisões judiciais ou políticas públicas.

Apesar de não ser de forma alguma incompatível com métodos de inferência causal, são distintos pelas restrições de ação a que estão sujeitos.

Em muitas situações reais o tempo disponível para se obter uma resposta razoável é menor, assim como a disponibilidade de dados. Por mais que idealmente leis e políticas públicas devessem ser pensadas no longo prazo, onde o fator tempo não é tão significativo, na prática muitas intervenções exigem uma avaliação de curto prazo sobre seu potencial impacto. Nestes casos o tempo necessário para coletar uma base de dados significativamente grande, e com uma estratégia de identificação adequada para as variáveis e efeitos que se quer medir, pode ser impossível.

No lugar da inferência causal – que certamente seria mais rigorosa –, métodos de CBAs servem para auxiliar na adoção de uma lógica de pressupostos, argumentação e aprofundamento das escolhas alternativas que dá um embasamento maior para a decisão pública. Por conta das limitações do método empírico empregado, o espaço para publicação destes trabalhos em periódicos é mais reduzido. E mesmo se houvesse, o fator limitante do tempo seria um empecilho de toda forma.

Assim, critérios de máximo rigor na estratégia de identificação são menos importantes que critérios satisfatórios à luz da literatura prévia sobre o tema, do conhecimento de especialistas da área, dos agentes específicos sobre os quais a política pública ou reforma jurídica vista, do estabelecimento razoável de valores para a tomada de decisão e tratamento explícito dos riscos envolvidos em diferentes escolhas.[30]

Para compensar a falta de revisão cega por pares, níveis elevados de transparência na exposição dos pressupostos, argumentos e bases de dados utilizadas são desejáveis e necessários. Menosprezar as Análises de Custo Benefício apenas por não terem o mesmo rigor estatístico e publicação em periódicos é arriscado, pois em geral, o substituto delas não é um artigo impecável em um periódico de fronteira, mas sim uma política pública, ou reforma jurídica irresponsável, sem qualquer interesse genuíno em avaliar seus custos e benefícios à luz das melhores evidências e razões *possíveis*. No lugar de uma análise razoável e transparente de custos e benefícios, ficamos reféns de análises sem qualquer transparência, e com custos e benefícios desconhecidos, dificultando a cobrança *ex-post*.

Estudiosos da AED têm muito a ganhar nesta área de pesquisa, assim como formuladores de políticas públicas, reformadores da legislação e juízes. O arcabouço metodológico da CBA auxilia o trabalho de encontrar melhores respostas e auxilia o trabalho acadêmico de avaliar as consequências das decisões tomadas.

Um breve comentário sobre outra área de aplicação de métodos empíricos em temas da AED, porém também com menor espaço nos periódicos científicos, é o emprego de novas técnicas estatísticas de previsão. A pesquisa econômica e do Direito em geral, e da AED em específico, nos periódicos não têm apresentado grande interesse em realizar previsões à revelia de considerações causais. No entanto, com técnicas modernas de aprendizado de máquina é possível que esse estado de coisas mude conforme novas ideias de como aliar técnicas de previsão com modelos causais e arcabouços teóricos.[31]

---

[30] BOARDMAN, Anthony; GREENBERG, David; VINING, Aidan *et al*. *Cost-Benefit Analysis*. 4ª ed. Boston: Pearson, 2010
; PEA - INSTITUTO DE PESQUISA ECONÔMICA APLICADA. *Avaliação de Políticas Públicas - Guia Prático de Análise Ex Ante*. Brasília: Casa Civil da Presidência da República, 2018, vol. 1; IPEA - INSTITUTO DE PESQUISA ECONÔMICA APLICADA. *Avaliação de Políticas Públicas - Guia Prático de Análise Ex Post*. Brasília: Casa Civil da Presidência da República, 2018, vol. 2.
[31] CHEN, Daniel L. *Machine Learning and the Rule of Law*. Law as Data, Santa Fe Institute Press, ed. M. Livermore and D. Rockmore, 2019; CHEN, Daniel L. "Judicial analytics and the

**Conclusões**

Neste capítulo meu objetivo foi apenas o de representar uma fotografia do estado de coisas de alguns problemas e métodos empíricos relevantes para a pesquisa em Análise Econômica do Direito e apontar caminhos em que a área está avançando. Para cada um dos tópicos aqui abordados há uma vasta literatura e o tratamento dado não foi mais que um "aperitivo". Nas bibliografias indicadas foi dada ampla preferência para livros e artigos que possibilitem o estudo mais aprofundado dos temas tratados, mas partindo do começo.

A consideração final mais importante que tenho a colocar é reiterar que considero o futuro da AED como programa de pesquisa profícuo estritamente dependente da criação de grupos de pesquisa grandes e multidisciplinares. A generalização de métodos estatísticos e computacionais para todas as áreas da ciência tende a diluir a primazia da AED como *a abordagem* mais centrada na análise empírica, positiva, de consequências. Manter a relevância da área, expandi-la e inová-la é possível, porém envolverá alavancar o conhecimento de fronteira em diferentes áreas da ciência.

Se hoje no Brasil vemos a Análise Econômica do Direito enfrentar resistência e dificuldade em avançar na pauta de que a avaliação rigorosa de consequências importam para o Direito, na fronteira do conhecimento mundial é possível que a dificuldade maior da AED nas próximas décadas será manter a importância metodológica das teorias econômicas e jurídicas clássicas ante a avalanche de novas áreas, métodos e estudos que avaliarão as consequências do Direito.

---

great transformation of American Law". *Artificial Intelligence and Law*, vol. 27, nº 1 (2019), pp. 15 – 42.

**Referências**